\documentstyle[preprint, revtex, eqsecnum]{aps}

\begin{document}
\columnsep0.1truecm
\draft
\preprint{ }
\begin{title}
Destructive quantum interference in spin tunneling problems
\end{title}
\author{Jan von Delft and Christopher L. Henley}
\begin{instit}
Laboratory of Atomic and Solid State Physics,
Cornell University, Ithaca, NY, 14853
\end{instit}
\receipt{July 27, 1992 / 9208011}
\begin{abstract}
In some spin tunneling problems, there are several different but
symmetry-related tunneling paths that connect the same initial and final
configurations. The
topological phase factors of the corresponding tunneling amplitudes
can lead to destructive interference between
the different paths, so that
the total tunneling amplitude is zero.
In the study of tunneling between different ground state
configurations of the Kagom\'{e}-lattice quantum Heisenberg
antiferromagnet, this occurs when the spin $s$ is half-odd-integer.
\end{abstract}


The problem of calculating the rate at which a quantum spin system
tunnels between its different low-energy states
 has been of interest in various different contexts  \cite{CG88}.
The tunneling amplitude is usually calculated by setting up
a coherent-spin-state path integral and analytically
continuing to imaginary time $(t \rightarrow - i \tau)$;
the leading contribution can be found using the method of steepest descent
\cite{Kla79}. The phase
of the tunneling amplitude depends on a topological phase \cite{footnote0}.

 In this Letter we point out that when there are several different
symmetry-related tunneling
paths connecting the same degenerate classical ground
states,   the topological phase can lead to destructive quantum
interference between their amplitudes,  and hence to
a total tunneling amplitude that is exactly zero.
The occurrence of such a cancellation has a geometric interpretation and
depends on the value of the spin $s$. We also present two examples:
The first involves a Hamiltonian with an $m$-fold symmetry axis.
The second concerns  tunneling
amplitudes between different degenerate ground state configurations of
a Kagom\'{e}-lattice quantum Heisenberg antiferromagnet; these
 amplitudes are zero if
 $s$ is half-odd-integer, but  non-zero if  $s$ is integer.

For extended quantum spin systems in one and two dimensions, qualitative
differences between integer and half-odd-integer spin $s$ have been reported
previously  \cite{Hal83}.
In these papers, quantum interference between topologically distinct
paths of a continuous unit vector field
{\boldmath{$\Omega$}}({\boldmath{$x$}},$t$)
leads to interesting $s$-dependencies.
Our examples show that similar effects occur for much simpler systems
that involve only a small number of individual spins.
\vspace{.1cm}

{\em The tunneling amplitude:}\/
In  tunneling problems, the customary object of study is the imaginary
time  transition amplitude
from an initial state $| i \rangle$ to a final state $|f \rangle$.
For a spin system, this can be written as a coherent-spin-state
 path integral  \cite{Kla79}:
\begin{equation}
\label{eq2}
U_{fi}  \, \equiv \,
\langle f | e^{- \widehat{{\cal H}} t / \hbar} | i \rangle = \int
\!{\cal D} \Omega
e^{- {\cal S} / \hbar} \; .
\end{equation}
where ${\cal S} = \int \!d \tau {\cal L} \!\!$ is the Euclidean action
and $\!{\cal D} \Omega$ is
the measure of the path integral.
For the special case of a single spin,  the Euclidean Lagrangian is
\begin{equation}
\label{eq1}
{\cal L}  = -i \hbar s \dot{\phi} (1 - \cos \theta) + {\cal H} (
\phi , \theta ) \; .
\end{equation}
 The coordinates $(\phi,\theta)$ label the coherent
spin state $| \phi,\theta \rangle$ for a particle with spin $s$,
and may be associated with a unit vector ${\mbox{\boldmath $n $}}$ in the
$(\phi, \theta) \equiv
\Omega$ direction. The dot on $\dot{\phi}$ means $\partial_{\tau}$.
The ``semi-classical'' Hamiltonian is the expectation value
 ${\cal H} \equiv \langle \phi,\theta | \widehat{{\cal H}}
| \phi,\theta \rangle$
of the operator $\widehat{{\cal H}} $. For simplicity, we shall
consider only the case
where $|i \rangle $ and $|f \rangle $ are ``classically degenerate ground
states" (in the sense that $\langle i | \widehat{{\cal H}}  | i
\rangle = \langle f |
\widehat{{\cal H}} | f \rangle$ are the smallest possible
expectation values of
$\widehat{{\cal H}}$), and are separated by an energy barrier.

The above path integral can be evaluated
by the method of steepest descent:
\begin{equation}
\label{eqL3}
U_{fi} \,  = \,  \sum_l {\cal N}^{(l)}
e^{- {{\cal S} }_o^{(l)} / \hbar } \, \equiv \, \sum_l U_{fi}^{(l)} \; ,
\end{equation}
\FL
\begin{equation}
\label{eqL3a}
{\rm where} \quad
{{\cal N}\,}^{(l)} = \int \! \! \!{\cal D} (\Omega -
\overline{\Omega}^{(l)}) e^{- (\delta^2
{\cal S}^{(l)} + \delta^3 {\cal S}^{(l)} \dots )  / \hbar} \; .
\end{equation}
Here  ${{\cal S} }_o^{(l)}$ is the action evaluated
along the $l$-th ``tunneling path", which is a solution
to the Lagrangian {\mbox equations} of motion and will be denoted by
overlined variables, e.g. $(\overline{\phi}^{(l)} (\tau),
\overline{\theta}^{(l)} (\tau) )$. The
index $l$ allows for the possibility of several different symmetry-related
tunneling paths. The prefactors ${{\cal N}\,}^{(l)}$ measure the effects of
fluctuations around the $l$-th tunneling path.
Tunneling problems are characterized by the fact  that
the coordinates in general acquire imaginary parts along the tunneling path
(else it is not possible to satisfy the requirement that the Hamiltonian
be a conserved quantity along the path). Consequently,
the various ${{\cal S} }_o^{(l)}$
can have non-zero imaginary parts. Quantum interference,
and possibly complete cancellation (so that $U_{fi} = 0$), can thus
occur between the amplitudes of the different paths.

{\em Geometric interpretation of phase:}\/
The conserved energy $E_o = {\cal H} (\overline{\Omega}^{(l)}) $
along the tunneling paths
that connect the degenerate states {\mbox{$ | i \rangle$}}\ and {\mbox{$ | f
\rangle$}}\
 may be set equal to zero without loss of generality.
Hence ${\cal S}_o \!\!^{(l)}$ is completely determined by the first term of
${\cal L}$.
This term has a well-known geometrical interpretation,
which we now discuss.

Let $\Omega ( \tau) = (\phi ( \tau), \theta ( \tau))$
be a purely real, {\em closed}\/, non-self-intersecting, smooth path in
spin space. The area on the unit sphere enclosed by this path is given by
\begin{equation}
\label{area}
A \, = \,  \int \! d \tau \, \dot{\phi} ( 1 - \cos \theta)  \; ,
\end{equation}
modulo $4 \pi$, depending on which of the two
oriented areas (the ``inside" or the ``outside" of the closed path)
one considers.

Now consider two (of the possibly many)
tunneling paths, say $\overline{\Omega}^{(l)} ( \tau)$ and
$\overline{\Omega}^{(l')} ( \tau)$,
not necessarily purely real, connecting {\mbox{$ | i \rangle$}}\ and
{\mbox{$ | f \rangle$}}.
Suppose that some symmetry of the Hamiltonian ensures that the absolute
values of the two tunneling
amplitudes $U_{fi}^{(l)}$ and $U_{fi}^{(l')}$ of eq.~(\ref{eqL3})
 are equal, i.e. that
\begin{equation}
\label{eq110.8}
{\rm Re} [ {\cal S}_o \!\!^{(l)} ] = {\rm Re} [{\cal S}_o \!\!^{(l')}] \; ,
\quad {\rm and} \quad
{{\cal N}\,}^{(l)} = {{\cal N}\,}^{(l')} \; .
\end{equation}
Intuitively speaking, this situation arises when
the local neighborhoods of the two tunneling paths are, for symmetry reasons,
identical for the two paths, so that
the local shapes and sizes of the barriers (which determine the ${\rm
Re} [{\cal S}_o \!\!^{(l)}]$'s) and the local fluctuations around the
tunneling paths
(which determine the ${{\cal N}\,}^{(l)}$'s), are identical.
Then $U_{fi}^{(l)}$ and $U_{fi}^{(l')}$ differ at most by a
phase, namely
\FL
\begin{eqnarray}
\nonumber
( {\cal S}_o \!\!^{(l)} - {\cal S}_o \!\!^{(l')} ) / \hbar & = & - i s \!
\left[
\int \!\! d \tau {\rm Re} [ \dot{\overline{\phi}}{}^{(l)} ] \left( 1 - {\rm
Re} [ \cos
\overline{\theta}{}^{(l)} ] \right) \right. \\ &- &
\label{eq111.3}
\left. \int \!\! d \tau {\rm Re} [ \dot{\overline{\phi}}{}^{(l')} ] \left(
1 - {\rm Re}
[ \cos \overline{\theta}{}^{(l')} ] \right) \right] \; .
\end{eqnarray}
Terms such as ${\rm Re}[\,\, ] {\rm Im}[\,\, ]$ do not appear
 due to eq.~(\ref{eq110.8}).
We have assumed that ${\rm Im}\,
[ \dot{\overline{\phi}}{}^{(l)}] {\rm Im}[ \cos \overline{\theta}{}^{(l)} ]
=  0$ (same for $l
\leftrightarrow l'$), for the
following reason: The conserved energy condition,
${\cal H} ( \overline{\Omega}^{(l)} (\tau) \, ) = 0$,
can be solved to find, for example,  $\overline{\theta}^{(l)}$ in terms of
$\overline{\phi}^{(l)}$. The dependent variable $\overline{\theta}^{(l)}$
will be a complex
function of the independent variable $\overline{\phi}^{(l)}$, which can be
taken to be real, so that ${\rm Im}[\phi^{(l)}] = 0$. A similar argument
works if one chooses to write $\overline{\phi}^{(l)}$ in terms of
$\overline{\theta}^{(l)}$ \cite{footnote01}.

Now, let $A_{ll'}$ be either one of the two oriented areas on the unit
sphere bounded by the loop formed by the two paths
 ${\rm Re}[ \overline{\Omega}^{(l)} ]$ and ${\rm Re} [
\overline{\Omega}^{(l')} ]$.  Then, by
eq.~(\ref{area}), the relative phase eq.~(\ref{eq111.3}) reduces to
\cite{footnote0}
\begin{equation}
\label{phasediff}
( {\cal S}_o \!\!^{(l)} - {\cal S}_o \!\!^{(l')} ) / \hbar = - i s A_{ll'}
\; .
\end{equation}
The $4 \pi$-ambiguity in
$A_{ll'}$ is irrelevant, since $\exp (-  i 4 \pi s)$ $= 1$ for any spin $s$.

If $s A_{ll'} $ is an odd multiple of $\pi$,
the amplitudes for the two paths interfere destructively, so that
$U_{fi}^{(l)} + U_{fi}^{(l')} = 0$.
In the simplest case where $\overline{\Omega}^{(l)}$ and
$\overline{\Omega}^{(l')} $
are the only {\mbox two} tunneling paths, this means that the {\em total}\/
amplitude $U_{fi}$ is zero.  Note that this result does not depend on the
detailed dynamics of the tunneling motion. In particular, it is not
necessary to know the imaginary parts of the tunneling paths
$\overline{\Omega}^{(l)}$ explicitly, as long as symmetry arguments can be
invoked to assert that the relations~(\ref{eq110.8}) hold. It may happen that
symmetry ensures that the prefactors\ ${{\cal N}\,}^{(l)}$ are
equal to {\em all}\/ orders of the steepest descent method, which is an
expansion in powers of $1/s$ (and not merely to the lowest order that is
usually employed). In this case the cancellation,
if it happens, is exact to all orders in $1/s$.

Clearly, the occurrence of destructive interference depends crucially on
the value of $s$. This could create  complications in
the analysis of the tunneling behavior of, for example,  a small
ferromagnetic particle
with effective spin $ns$ (as has been considered in  \cite[c]{CG88}),
 where $n$ is the number of spins in the
particle, since the effective spin is then $n$-dependent.

Turning on an external magnetic field can in general
affect the occurrence of destructive interference by changing
the initial conditions (i.e.\ {\mbox{$ | i \rangle$}}\ and {\mbox{$ | f
\rangle$}}) and
the area $A_{ll'}$ enclosed between different tunneling paths, and by
destroying the symmetry that ensures that the absolute values of the
amplitudes for all tunneling paths are the same.
\vspace{.2cm}

We now present two examples where destructive interference does occur.

{\em Hamiltonian with $m$-fold symmetry axis:}\/
Consider a single spin $s$. Suppose that the Hamiltonian has an easy axis
(say the $z$-axis, $\theta = 0$), around which it has $m$-fold rotational
symmetry:
${\cal H} (\phi, \theta) \equiv {\cal H} (\phi + 2 \pi / m,
\theta)$ for all $\phi$.
Suppose that ${\mbox{$ | i \rangle$}} = | \theta = 0 \rangle $ and $
{\mbox{$ | f \rangle$}} = | \theta = \pi
\rangle $ correspond to the two degenerate classical ground state
configurations. Clearly, if $ \overline{\Omega}^{(0)} =     (
\overline{\phi}^{(0)},  \overline{\theta}^{(0)} )$  is a tunneling path
from {\mbox{$ | i \rangle$}}\ to {\mbox{$ | f \rangle$}}, so are the paths
$\overline{\Omega}^{(l)} = ( \overline{\phi}^{(0)} + 2 \pi l / m ,
\overline{\theta}^{(0)} ) $,
for $l= 0, \dots , m-1$ (see fig.~\ref{fig1}).
By symmetry, all ${\rm Re} [ {\cal S}_o \!\!^{(l)} ]$ are equal to each
other, as are all $\, {{\cal N}\,}^{(l)}$. Furthermore,
\begin{equation}
\label{eq112.7}
({\cal S}_o \!\!^{(0)} - {\cal S}_o \!\!^{(l)} ) / \hbar \, = \, - i s
A_{0l} \, =  \, - i s 4 \pi l / m \; ,
\end{equation}
because the area on the unit sphere enclosed between the real parts of any
two neighboring paths is, by symmetry, necessarily equal to $4 \pi / m$.
The total amplitude is thus:
\FL
\[
U_{fi} = {{\cal N}\,}^{(0)} e^{- {\cal S}_o \!\!^{(0)}} \!
\sum_{l = 0}^{m-1} e^{ i 4 \pi l s / m}
 =  \delta_{2s, km} \, m \, {{\cal N}\,}^{(0)} e^{- {\cal S}_o
 \!\!^{(0)} / \hbar} ,
\]
where the $\delta$-function is non-zero only if $2s$ is an integer
multiple of $m$.

A realization of the above scenario, with $n = 3$, is afforded by a
Hamiltonian with cubic anisotropy and a small additional uniaxial
anisotropy along the $[111]$ direction (thus 3-fold symmetry around $[111]$):
\FL
\begin{equation}
\label{eq113.1}
{\cal H} = J_1 \left( n_x^4 + n_y^4 + n_z^4 - {\textstyle \frac{1}{3}}
\right)
+ J_2 \left( 1 - ( {\mbox{\boldmath $n $}} \cdot {\mbox{\boldmath $d $}})^2
\right) \; ,
\end{equation}
where $n_x$, $n_y$ and $n_z$ are the direction cosines of the
unit vector ${\mbox{\boldmath $n $}}$ in the $(\phi, \theta)$ direction, and
$J_1 \gg J_2 > 0$.
The unit vectors ${\mbox{\boldmath $d $}} \equiv \frac{1}{\sqrt{3}} (1,1,1)$
and $-{\mbox{\boldmath $d $}}$ define the two classical ground-state
configurations  \cite{footnote1b}.
\vspace{.1cm}

{\em Spin tunneling in the Kagom\'e lattice:}\/
Our second example concerns spin tunneling events in a $2D$
 quantum Hei\-senberg nearest-neighbor antiferromagnet on a Kagom\'e lattice
 (fig.~\ref{fig3})
\cite{HKB92}. The Hamiltonian is taken to be
\begin{equation}
\label{eq?}
{\cal H} = s^2 J \sum_{ \langle i,j \rangle}
{\mbox{\boldmath $n $}}_i \! \cdot \!  {\mbox{\boldmath  $n $}}_j \; ,
 \qquad (J > 0) \; ,
\end{equation}
where $s {\mbox{\boldmath $n $}}_i \equiv \langle \phi_i, \theta_i |
\widehat{{\mbox{\boldmath $s $}}}_i | \phi_i,
\theta_i \rangle $ is  the ``classical'' spin
(see fig.~\ref{fig2}a). Any configuration in which
the spins on {\em each}\/ triangle minimize their energy by
assuming a coplanar configuration, with relative angles of $120^{\circ}$,
(see fig.~\ref{fig2}b), is a classical ground state.
Therefore there are macroscopically many degenerate classical ground states.
Generally fluctuations, both quantum and thermal, are expected to lift
some of the degeneracies, thereby inducing magnetic
ordering by selecting particular configurations
(``order from disorder"). For example, spin wave expansions about various
ground state configurations have shown
that maximally coplanar configurations, in which {\em all}\/ spins in the
lattice lie in the same plane (let this ``reference plane'' define $\phi = 0$
and $\phi = \pi$), have lower zero-point energies than any other
configurations \cite[b,d]{HKB92}.

On the other hand, tunneling  between
different degenerate ground state configurations competes with ``order from
 disorder'' selection effects, because it tends to drive the system into a
superposition of degenerate states, rather than selecting a particular one.
As the simplest example of a tunneling event on the Kagom\'e lattice,
we consider the so-called ``weathervane mode'' (see fig.~\ref{fig3}): the six
spins of an $ABABAB$-hexagon in one maximally coplanar ground state,
{\mbox{$ | i \rangle$}},
rotate synchronously by $180^{\circ}$ around the {\mbox{\boldmath $z
$}}-axis (defined by the
$C$-spins), to end up as a $BABABA$-hexagon in another maximally coplanar
ground state, {\mbox{$ | f \rangle$}}, while all other spins remain fixed.
Due to the
above-mentioned spin wave selection effects, there is a ``coplanarity
barrier'', say $J_b (\phi)$, to this type of motion, which, by reflection
symmetry in the reference plane, obeys $J_b (\phi) = J_b (- \phi)$.

To study the hexagon tunneling event, we consider the following
Euclidean Lagrangian:
\FL
\begin{eqnarray}
\nonumber
{\cal L} \!&= &\! \sum_{j = 1}^6 \! - i \hbar s \dot{\phi}_j (1 - \cos
\theta_j)  + J_b \! \sum_{j = 1}^6 \sin^2 \!
\textstyle{ \frac{1}{2}} ( \phi_j + \phi_{j+1} + \pi)  \\
& & + \;  s^2 J \sum_{j = 1}^6 \bigl[ {\mbox{\boldmath $n $}}_j \! \cdot \!
{\mbox{\boldmath $n $}}_{j+1} + {\mbox{\boldmath $n $}}_j \! \cdot \!
{\mbox{\boldmath $z $}} + {\mbox{\boldmath $n $}}_{j+1} \! \cdot \!
{\mbox{\boldmath $z $}} + {\textstyle \frac{3}{2}} \bigr] .
\label{eq2.5}
\end{eqnarray}
 The index $j$ is defined modulo 6. We take $J \gg J_b >0$.
The $s^2 J {\mbox{\boldmath $n $}}_j \! \cdot \! {\mbox{\boldmath $z $}}$
terms (interaction with $C$-spins) and
 the phenomenological $\sin^2 \! \phi$ coplanarity
 barrier \cite{footnote1} are assumed to be the only ways the other spins
in the lattice, which are assumed to remain fixed,
 influence the six spins on the hexagon.

To minimize the cost of the dominant $s^2 J$ term, the hexagon spins are
expected to rotate collectively, maintaining mutual near-coplanarity.
Indeed, it can be shown \cite{vDH92}
that the hexagon tunneling problem can be mapped onto
a simple model problem, defined by the Lagrangian
\FL
\[
{\cal L} \!\! = - i \hbar 6s \dot{ \phi}
(1 - \cos \theta) + 12 s^2 J (\cos \theta - {\textstyle \frac{1}{2}})^2
 + 6 J_b \sin^2 \phi \, ,
\]
 involving only a single (collective) spin degree of freedom with
an effective spin of $6s$. For present
purposes, however, the following observations suffice (discussed in detail
in \cite{vDH92}):
Due to the reflection symmetry about the reference plane,
there are two possible tunneling paths, to be denoted by
$(\overline{\phi}^{\pm}, \overline{\theta}^{\pm})$; they differ
from each other only in the {\em direction}\/ of the
$\overline{\phi}$-rotations and satisfy $\overline{\phi}_j^{-} (\tau) = -
\overline{\phi}_j^{+} (\tau) $, for  $j =1, \dots, 6$.
Reflection symmetry ensures that ${\rm Re}[{\cal S}_o \!\!^+] =
{\rm Re}[{\cal S}_o \!\!^-] $ and ${{\cal N}\,}^+ = {{\cal N}\,}^-$.
Along both tunneling paths, every
$\overline{\phi}^{\pm}_j$ is pure\-ly real. To satisfy ${\cal H} =
0$ during the tunneling
event, each $\theta_j$
develops a time-dependent imaginary part (which vanishes
in {\mbox{$ | i \rangle$}}\ and {\mbox{$ | f \rangle$}}), but ${\rm Re}
[\cos \overline{\theta}_j]$
maintains the value it has in {\mbox{$ | i \rangle$}}\ and {\mbox{$
| f \rangle$}}, namely ${\rm Re} [\cos
\overline{\theta}_j] = - \textstyle{ \frac{1}{2}}$ $j =1, \dots, 6$.

Thus, for each of the six spins, the real part of the tunneling path is a
contour of constant ${\rm Re}[\cos \theta_j] = - \textstyle{ \frac{1}{2}}$,
with ${\rm Re}
 [\overline{\phi}^+_j] $ (or ${\rm Re} [\overline{\phi}^-_j] $)
changing from 0 to $\pi$ (or $- \pi$) for $(+)$- or $(-)$-paths.
For each spin, the area enclosed between the $(+)$- and  $(-)$-paths
is thus equal to $\pi$, giving for the six spins a
 total phase difference of $i 6 \pi s$ between the amplitudes for an
$(+)$- or $(-)$-event.
It follows that the total tunneling amplitude be\-comes
\begin{eqnarray}
\label{eq114.6}
U_{fi} & = & {{\cal N}\,}^+ e^{- {\cal S}_o \!\!^+ / \hbar} ( 1 + e^{i
 6 \pi s}) \\ \nonumber &=&
\left\{
\begin{array}{ll}
2 {{\cal N}\,}^{+} e^{- {\cal S}_o \!\!^{+} / \hbar} \quad & {\mbox{if
$s$ is integer}}  \; , \\
0 & {\mbox{if $s$ is half-odd-integer}}  \; .
\end{array} \right.
\end{eqnarray}

Similarly, consider any larger closed ``loop" of alternating $A$-
and $B$-spins within a ground-state configuration. It can be proven that any
such loop contains $4n + 2$ spins ($n$ is some  integer) \cite{vDH92}.
 Again one can study the tunneling between two configurations
that only differ by  $\phi_l \rightarrow \phi_l + \pi$
(i.e. $A \leftrightarrow B$)
for each spin on the loop. The relevant phase between $(+)$- and $(-)$-
paths will be $ i \pi  (4n +
2) s $, and for half-odd-integer $s$, destructive interference again occurs.

The above results imply that in that subset of parameter space where
``order from disorder'' selection effects and the competing
tunneling effects that favor more disorder are more or less equally
important, one might expect interesting integer versus half-odd-integer $s$
effects, reminiscent of those found in 1D antiferromagnetic spin
chains \cite[a]{Hal83}. It would therefore be of interest to find
integer-spin realizations of the Kagom\'e lattice antiferromagnet,
since $s$ is half-odd-integer for
the two experimental realizations of the Kagom\'{e} lattice that have
been proposed: the  magnetoplumbite $SrCr_{8-x}Ga_{4+x}O_{19}$
($s= \frac{3}{2}$)
 and the second layer of $He^3$ atoms on a graphite substrate ($s =
\frac{1}{2}$) \cite{BAE90}.

Integer versus half-odd-integer $s$ effects might
also make their appearance in exact diagonalization studies of
finite size systems with discrete degeneracies, in the analysis of
which spin tunneling methods should be very useful. The
simplest realization of such a discrete degeneracy is the $J_1 - J_2$
square lattice antiferromagnet for $J_2 > J_1/2$, which shows a discrete
degeneracy between antiferromagnetic ordering vectors $(1,0)$ and $(0,1)$
\cite{PGBD91}.

The vanishing of tunneling amplitudes obviously implies an exact
ground state degeneracy in the semi-classical picture. Sometimes, this
degeneracy can be shown to exist for {\em all}\/ eigenstates of the
system, on a purely quantum-mechanical level.
 For example, consider the following toy
model for a quantum spin (independently suggested to us by V. Elser):
 ${\cal H} = - \widehat{S}_z^2 - a \widehat{S}_x^2$. Here
$a \ll 1$, and $\widehat{S}_z$ and $\widehat{S}_x $ are spin operators.
The tunneling amplitude between the
two classical ground states, ${\mbox{\boldmath $n $}} = \pm {\mbox{\boldmath
$z $}}$ along the two tunneling
paths $\overline{\theta}^{\pm}: 0 \rightarrow \pi$, $\overline{\phi}^+ = 0$,
  $\overline{\phi}^- = \pi$,
is zero for half-odd-integer $s$ (the phase difference is $2 \pi s$).
On the other hand, ${\cal H}$ has time-reversal symmetry and hence all
its eigenstates display a two-fold Kramers degeneracy for half-odd-integer
$s$. It would be interesting to investigate more generally
under what circumstances
vanishing semi-classical tunneling amplitudes also imply exact
quantum-mechanical degeneracies for all eigenstates.

In conclusion, we have shown that the topological phase factor
occurring in spin tunneling amplitudes can have quite striking effects.
If there are several symmetry-related tunneling paths whose individual
tunneling amplitudes have the same absolute value, their topological
phases can lead to destructive quantum interference between the paths and
a total tunneling amplitude that is zero. The conditions under which
this occurs can be interpreted geometrically in terms of the areas on the
unit sphere enclosed between the real parts of the various tunneling
paths. In quantum antiferromagnets,  interesting
integer versus
half-odd-integer $s$ effects can result from the competition between
``order from disorder'' selection and tunneling effects.

Helpful discussions with A. Auerbach,  D.P.
DiVincenzo, V. Elser, A. Garg, M. Kvale and  S. Sachdev
are gratefully acknowledged. Related preliminary calculations were done
by Q. Sheng.  Part of this work was supported by N.S.F. grant DMR--9045787.

\vspace{2cm}

\figure{Three equivalent tunneling paths for a Hamiltonian with 3-fold
rotation symmetry on a ``wrapped-open'' unit sphere. \label{fig1}}
\vspace{2cm}

\figure{An $ABABAB$ hexagon of spins (on the left) in a coplanar ground state
configuration.  \label{fig3}}

\vspace{2cm}

\figure{(a) The unit vector ${\mbox{\boldmath $n $}} = (\phi, \theta)$. (b)
 Type $A$, $B$
and $C$ spins on a triangle of the Kagom\'e lattice.  \label{fig2}}


\begin{thebibliography}{WWW99}
\bibitem{footnote0}
This relative phase may be thought of as a Berry phase \cite[b]{Hal83}.
\bibitem{footnote01} Examples of this procedure may be found in
\cite[c]{CG88}.
\bibitem{footnote1} For a different class of tunneling events, a $| \sin
\phi |$ coplanarity barrier has been found by \cite[b]{HKB92}. However, our
argument does not depend on the precise form of the barrier, as long as
$J_b (\phi) = J_b (- \phi)$.
\end{thebibliography}
\end{document}